\documentclass[twocolumn,aps,nofootinbib,showpacs,preprintnumbers,amsmath,amssymb]{revtex4}
\usepackage{graphicx}
\usepackage{dcolumn}
\usepackage{bm}
\usepackage{epsfig}
\usepackage{epsf}
\usepackage{amssymb}
\usepackage{amsmath}

\newcommand{\A}{\mathcal{A}}
\newcommand{\tA}{{\widetilde {\mathcal{A}}}}
\newcommand{\tlt}{{\widetilde t}}
\newcommand{\tlA}{\mathfrak A}
\newcommand{\be}{\begin{equation}}
\newcommand{\ee}{\end{equation}}
\newcommand{\bep}{\begin{equation*}}
\newcommand{\eep}{\end{equation*}}
\newcommand{\ba}{\begin{eqnarray}}
\newcommand{\ea}{\end{eqnarray}}
\newcommand{\bap}{\begin{eqnarray*}}
\newcommand{\eap}{\end{eqnarray*}}
\newcommand{\bMS}{{\overline {\rm MS}}}
\newcommand{\bL}{{\overline \Lambda}}
\newcommand{\bM}{{\overline M}}

\begin{document}
\preprint{USM-TH-227}

\title{Analytic QCD -- a short review}

\author{Gorazd Cveti\v{c}$^{a,b}$}
  \email{gorazd.cvetic@usm.cl}
\author{Cristi\'an Valenzuela$^c$}
  \email{cvalenzuela@fis.puc.cl}
\affiliation{$^a$\,Center of Subatomic Studies and\\
$^b$\,Dept.~of Physics, Univ.~T\'ecnica
F.~Santa Mar\'{\i}a, Valpara\'{\i}so, Chile\\
$^c$\,Dept.~of Physics, Pontif.~Univ.~Cat\'olica de Chile, Santiago 22, Chile}

\date{\today}

\begin{abstract}
Analytic versions of QCD are those whose coupling $\alpha_s(Q^2)$ does 
not have the unphysical Landau singularities on the space-like axis
($-q^2=Q^2 > 0$). The coupling is analytic in the entire complex plane 
except the time-like axis ($Q^2 < 0$). Such couplings are thus suitable 
for application of perturbative methods down to energies of order GeV. 
We present a short review of the activity in the area which started
with a seminal paper of Shirkov and Solovtsov ten years ago.
Several models for analytic QCD coupling are presented. 
Strengths and weaknesses of some of these models are pointed out.
Further, for such analytic couplings, constructions of the 
corresponding higher order analytic couplings (the analogs of the 
higher powers of the perturbative coupling) are outlined, and an
approach based on the renormalization group 
considerations is singled out. Methods of evaluation of the 
leading-twist part of space-like observables in such analytic frameworks 
are described. Such methods are applicable also to the inclusive 
time-like observables. Two analytic models are outlined which respect 
the ITEP Operator Product Expansion philosophy, and thus allow for an 
evaluation of higher-twist contributions to observables.
\end{abstract}
\pacs{12.38.Cy, 12.38.Aw,12.40.Vv}

\maketitle

\section{Introduction}
\label{sec:intro}

Perturbative QCD calculations involve coupling 
$a(Q^2) \equiv \alpha_s(Q^2)/\pi$
which has Landau singularities (poles, cuts)
on the space-like semiaxis
$0 \leq Q^2 \leq \Lambda^2$  ($q^2 \equiv - Q^2$). 
These lead to Landau singularities for the evaluated 
space-like observables ${\cal D}(Q^2)$ at low 
$Q^2 \stackrel{<}{\sim} \Lambda^2$. The existence of
such singularities is in contradiction with the
general principles of the local quantum field
theories \cite{BSh}. Further, lattice simulations
\cite{ls} confirm that such singularities
are not present in $a(Q^2)$. 

An analytized coupling $\A_1(Q^2)$, 
which agrees with the perturbative $a(Q^2)$
at ${Q^2 \to \infty}$ and is analytic
in the Euclidean part 
of the $Q^2$-plane ($Q^2 \ \epsilon \ {\cal C}$, $Q^2 \not \leq 0$), 
addresses this problem, and has been
constructed by Shirkov and Solovtsov about ten years ago \cite{ShS}.

Several other analytic QCD (anQCD) models for ${\A_1(Q^2)}$ 
can be constructed, possibly satisfying 
certain additional constraints at low and/or at high $Q^2$.  

Another problem is the analytization of higher power terms
${a^n} \mapsto \A_n$ in the truncated perturbation series (TPS) 
for ${\cal D}(Q^2)$. 
Also here, several possibilities appear. 

Application of the Operator Product Expansion (OPE) approach,
in the ITEP sense, to inclusive space-like observables 
appears to make sense only in a restricted class of such anQCD 
models. 

This is a short and incomplete review of the
activity in the area; relatively large space 
is given to the work of the review's authors.
For an earlier and more extensive 
review, see e.~g.~Ref.~\cite{Prosperi:2006hx}.  

Section \ref{sec:analyt} contains general aspects of
analytization of the {Euclidean coupling} 
${a(Q^2) \mapsto \A_1(Q^2)}$, and 
the definition of the {time-like (Minkowskian) coupling} 
${\tlA_1(s)}$. Further, in Sec.~\ref{sec:analyt} we review the
minimal analytization (MA) procedure developed by
Shirkov and Solovtsov
\cite{ShS}, and a variant thereof developed 
by Nesterenko \cite{Neste2000}.
In Sec.~\ref{sec:models} we present various
approaches of going beyond the MA procedure, i.e.,
various models for ${\tlA_1(s)}$,
and thus for ${\A_1(Q^2)}$ 
\cite{Alekseev:2005he,Srivastava:2001ts,Webber:1998um,Neste2005,Cvetic:2006mk,Cvetic:2006gc}.
In Sec.~\ref{sec:hipow}, analytization procedures for the higher powers
${ a^n(Q^2) \mapsto  \A_n(Q^2)}$ in MA model are presented
\cite{Milton:1997mi,Milton:2000fi,Sh}, and an
alternative approach which is applicable to any model of 
analytic ${\A_1(Q^2)}$ \cite{Cvetic:2006mk,Cvetic:2006gc} is presented.
In Sec.~\ref{sec:nonint}, an analytization of noninteger
powers $a^{\nu}(Q^2)$ is outlined \cite{Bakulev}. In Sec.~\ref{sec:eval},
methods of evaluations of space-like and of inclusive time-like 
observables in models with analytic ${\A_1(Q^2)}$ are described,
and some numerical results are presented for semihadronic $\tau$ 
decay rate ratio ${r_{\tau}}$, Adler function $d_v(Q^2)$ and 
Bjorken polarized sum rule (BjPSR) ${d_b(Q^2)}$
\cite{Milton:1997mi,Milton:2000fi,Sh,Shirkov:2006gv,Cvetic:2006mk,Cvetic:2006gc}.
In Sec.~\ref{sec:itep}, two sets of models are presented 
\cite{Raczka,Cvetic:2007ad} whose
analytic couplings ${\A_1(Q^2)}$ preserve the
OPE-ITEP philosophy, i.e., at high $Q^2$ they fulfill:
${|\A_1(Q^2)-a(Q^2)| < (\Lambda^2/Q^2)^k}$ for any
$k \ \epsilon \ {\cal N}$. Section \ref{sec:summ} contains
a summary of the presented themes.

\section{Analytization ${a(Q^2)} \mapsto {\A_1(Q^2)}$}
\label{sec:analyt}

In perturbative QCD (pQCD), the beta function is written as
a truncated perturbation series (TPS) of coupling $a$.
Therefore, the renormalization group equation (RGE) for ${a}(Q^2)$
has the form
\ba
\frac{\partial {a}(\ln Q^2; {\beta_2}, \ldots)}
{\partial \ln Q^2} 
& = &
- \sum_{j=2}^{j_{\rm max}} {\beta_{j-2}} \: 
{a}^j (\ln Q^2; {\beta_2}, 
\ldots) .
\label{pRGE}
\ea
The first two coefficients
[${\beta_0}= (1/4)(11-2 {n_f}/3)$,
${\beta_1}=(1/16)(102-38 {n_f}/3)$]
are scheme-independent in mass-independent schemes.
The other coefficients $({\beta_2, {\beta_3}, \ldots})$
characterize the renormalization scheme (RSch).
The solution of perturbative RGE (\ref{pRGE}) can be written
in the form
\be
{a}(Q^2)=
\sum_{k=1}^{\infty} \sum_{\ell=0}^{k-1} K_{k \ell} \:
\frac{(\ln {L})^{\ell}}{{L}^k},
\label{apt}
\ee
where ${L}=\ln(Q^2/{\Lambda^2})$
and $K_{k \ell}$ are constants depending on $\beta_j$'s.
In ${\bMS}$: ${\Lambda} = {\bL} \sim 10^{-1}$ GeV. 

\begin{figure}[htb]
\begin{minipage}[b]{.49\linewidth}
 \centering\epsfig{file=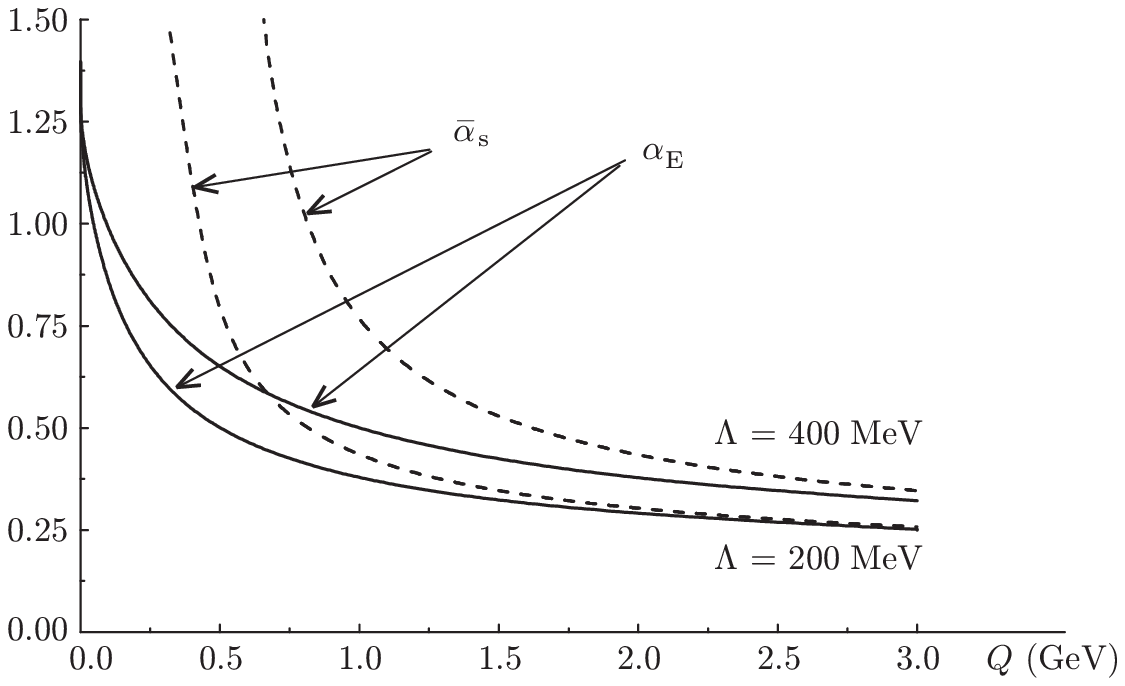,width=\linewidth,height=4.cm}
\end{minipage}
\begin{minipage}[b]{.49\linewidth}
 \centering\epsfig{file=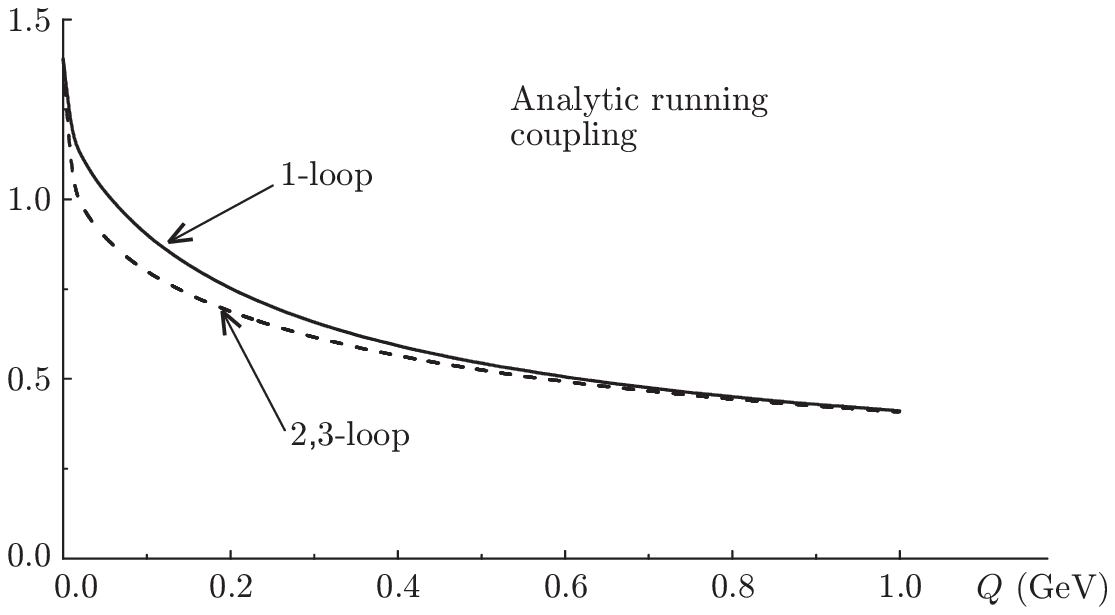,width=\linewidth,height=4.cm}
\end{minipage}
\vspace{-0.3cm}
\caption{\footnotesize Left: one-loop MA $\alpha_{\rm E}(Q)=\pi {\A_1}(Q^2)$ and 
its one-loop perturbative counterpart ${\overline \alpha}_s(Q^2)$
in ${\bMS}$, for ${n_f}=3$ and 
${\Lambda} = {\bL} = 0.2$ and $0.4$ GeV. Right: stability of the MA $\alpha_{\rm E}(Q)=\pi {\A_1}(Q^2)$
under the loop-level increase. Both figures from: Shirkov and Solovtsov, 1997 \cite{ShS}.}
\label{shir1shir2}
\end{figure}

\begin{figure}[htb] 
\centering
\epsfig{file=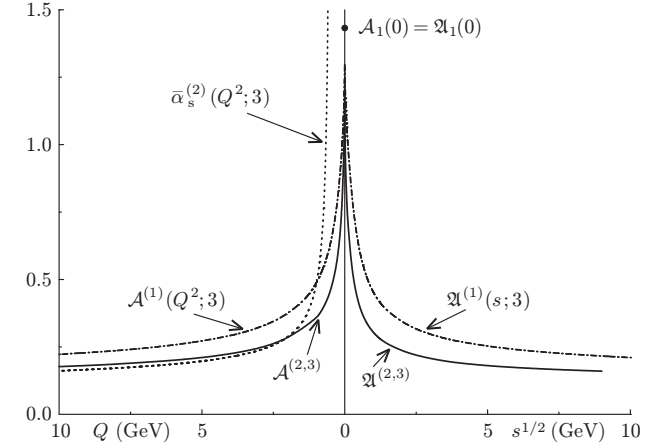,width=8.2cm}
\vspace{-0.3cm}
\caption{\footnotesize
The MA time-like and space-like
couplings ${\tlA_1}(s^{1/2})$ and 
${\A_1}(Q)$ at 1-loop, 2-loop (3-loop) level; 
in ${\bMS}$
for ${n_f}=3$ and ${\bL} = 0.35$ GeV
[${\tlA_1}$ and $\A_1$ in figure are
$\pi {\tlA_1}$ and $\pi \A_1$ in our normalization convention]. 
Figure from: Shirkov and Solovtsov, 2006 \cite{Shirkov:2006gv}.}
\label{shir3a}
 \end{figure}

\begin{figure}[htb]
\begin{minipage}[b]{.49\linewidth}
 \centering\epsfig{file=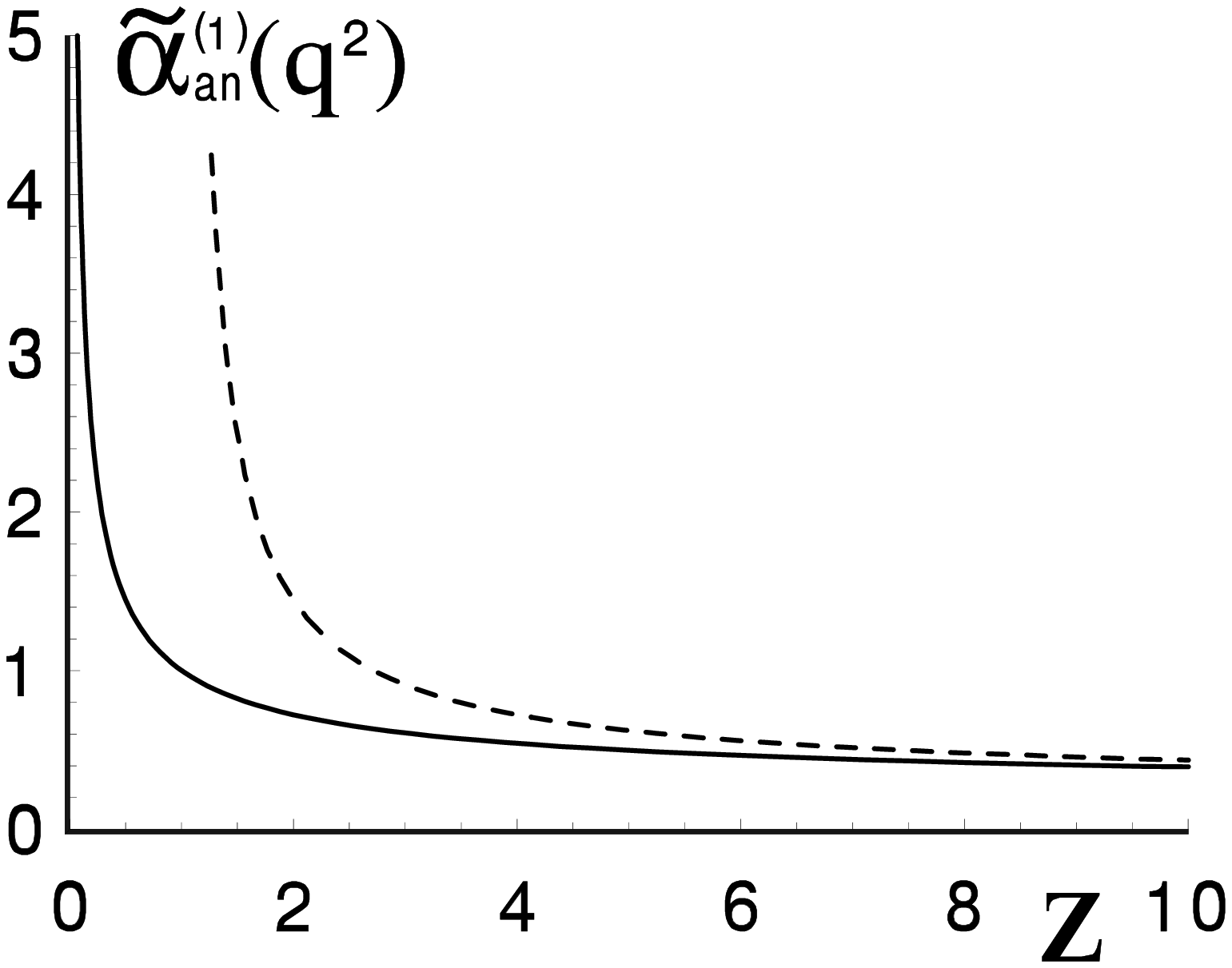,width=\linewidth,height=4.cm}
\end{minipage}
\begin{minipage}[b]{.49\linewidth}
 \centering\epsfig{file=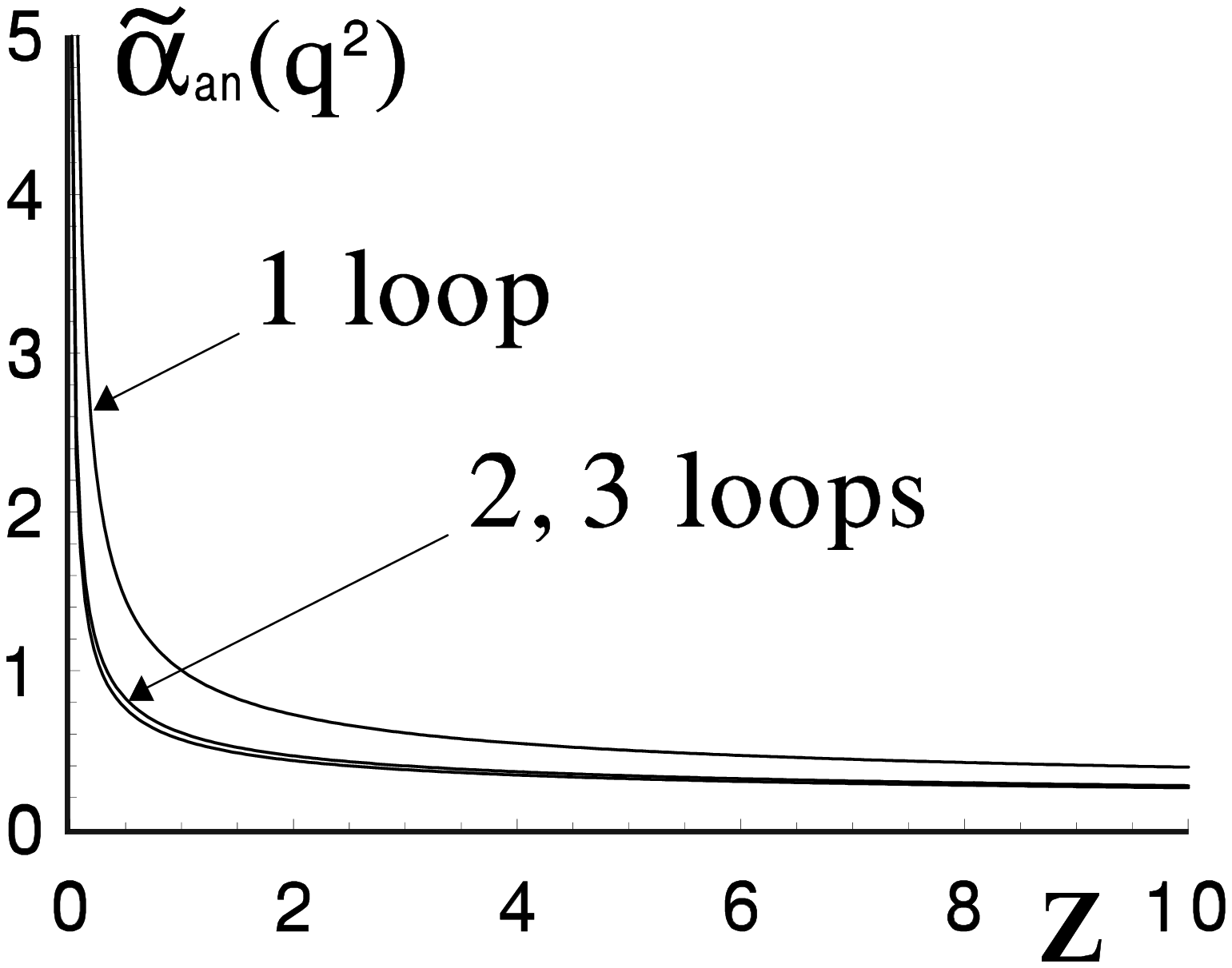,width=\linewidth,height=4.cm}
\end{minipage}
\vspace{-0.3cm}
\caption{\footnotesize
Left: one-loop ${{\overline {\rm MA}}}$
${{\widetilde \alpha}}_{\rm an}(Q)=
{\beta_0} {\A_1}(Q^2)$ and 
its one-loop perturbative counterpart,
as a function of $Z = Q^2/{\Lambda}^2$
(Figure from: Nesterenko, 2000 \cite{Neste2000}).
Right: stability of the  ${{\overline {\rm MA}}}$
${{\widetilde \alpha}}_{\rm an}(Q)=
{\beta_0} {\A_1}(Q^2)$
under the loop-level increase,
as a function of $Z = Q^2/{\Lambda}^2$
(Figure from: Nesterenko, 2001 \cite{Neste2001}).}
\label{Nest20002001}
\end{figure}

\begin{figure}[htb] 
\centering
\epsfig{file=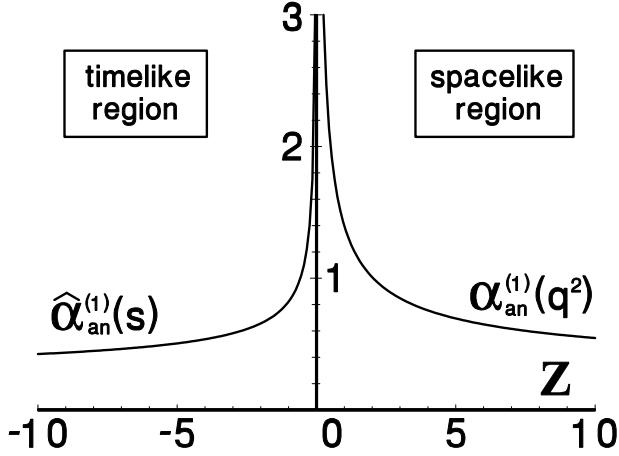,width=8.2cm}
\vspace{-0.3cm}
\caption{\footnotesize
One-loop time-like and space-like 
${{\overline {\rm MA}}}$ couplings
${{\hat \alpha}}_{\rm an}(s)=
\pi {\tlA_1}(s)$ and
${{\alpha}}_{\rm an}(Q^2)= \pi {\A_1}(Q^2)$ 
as a function of $Z = -s/{\Lambda}^2$ or 
$Z = Q^2/{\Lambda}^2$, respectively. 
Figure from: Nesterenko, 2003 \cite{Neste2003}.}
\label{Nest2003}
\end{figure}

The pQCD coupling ${a}(Q^2)$ is nonanalytic on 
${-\infty} < Q^2 \leq {\bL^2}$.
Application of the Cauchy theorem gives the dispersion relation
\begin{equation}
{a}(Q^2) = \frac{1}{\pi} \int_{\sigma= - {\Lambda}^2 - \eta}^{\infty}
\frac{d \sigma {\rho^{\rm {(pt)}}_1}(\sigma) }{(\sigma + Q^2)},
   \quad (\eta \to 0),
\label{aptdisp}
\end{equation}
where ${\rho^{\rm {(pt)}}_1}(\sigma)$ is the (pQCD) 
{discontinuity function} of ${a}$
along the cut axis in the $Q^2$-plane:
${\rho^{\rm {(pt)}}_1}(\sigma)= {\rm Im} {a}(-\sigma - i \epsilon)$.
The MA procedure of Shirkov and Solovtsov \cite{ShS} removes the
pQCD contribution of the unphysical cut $0 < - \sigma \leq \Lambda^2$,
keeping the discontinuity elsewhere unchanged 
(``minimal analytization'' of $a$) 
\begin{equation}
{\A^{\rm {(MA)}}_1}(Q^2) = \frac{1}{\pi} \int_{\sigma= 0}^{\infty}
\frac{d \sigma {\rho^{\rm {(pt)}}_1}(\sigma) }{(\sigma + Q^2)} \ .
\label{MAA1disp}
\end{equation}
In general:
\begin{equation}
{\A_1}(Q^2) = \frac{1}{\pi} \int_{\sigma= 0}^{\infty}
\frac{d \sigma {\rho_1}(\sigma) }{(\sigma + Q^2)} \ ,
\label{A1disp}
\end{equation}
where 
${\rho_1}(\sigma) = {\rm Im} {\A_1}(-\sigma - i \epsilon)$ .
Relation (\ref{A1disp}) defines an analytic coupling 
in the entire Euclidean complex $Q^2$-plane,
i.e., excluding the time-like semiaxis
$-s = Q^2 \leq 0$. On this semi-axis, it is convenient to
define the {time-like (Minkowskian) coupling} ${\tlA_1}(s)$
\cite{Milton:1997mi,Milton:2000fi,Sh}
\begin{equation}
{\tlA_1}(s) = \frac{i}{2 \pi} 
\int_{- s + i \epsilon}^{-s - i \epsilon} 
\frac{d \sigma^{\prime}}{\sigma^{\prime}}
{\A_1}(\sigma^{\prime}) \ .
\label{tlA1A1}
\end{equation}
The following relations hold between ${\A_1}$, ${\tlA_1}$
and ${\rho_1}$:
\begin{eqnarray}
{\tlA_1}(s) &=& \frac{1}{\pi} 
\int_s^{\infty} \frac{d \sigma}{\sigma}
{{\rho}_1}(\sigma) \ ,
\label{tlA1}
\\
{\A_1}(Q^2) & = & Q^2 \int_0^{\infty} 
\frac{ ds {{\tlA}_1}(s) }{(s + Q^2)^2} \ ,
\label{A1tlA1}
\\
\frac{d}{d \ln \sigma} {{\tlA}_1}(\sigma) &=&
- \frac{1}{\pi} {{\rho}_1}(\sigma) \ .
\label{tlA1rho1}
\end{eqnarray} 
The MA is equivalent to the 
minimal analytization of the
TPS form of the ${\beta}({a})=
\partial {a}(Q^2)/\partial \ln Q^2$ function
\cite{Magradze}
\ba
\frac{\partial {\A_1}^{\rm {(MA)}}(\ln Q^2; {\beta_2}, \ldots)}
{\partial \ln Q^2} 
& = &
\frac{1}{\pi} \int_{\sigma=0}^{\infty}
\frac{d \sigma {\rho^{\rm {(pt)}}_{\beta}}(\sigma) }{(\sigma + Q^2)},
\label{pRGEMagr}
\ea
where ${\rho^{\rm {(pt)}}_{\beta}}(\sigma)= 
{\rm Im} { \beta(a)}(-\sigma - i \epsilon)$,
and 
\ba
{ \beta}({a}) & = &
- \sum_{j=2}^{j_{\rm max}} {\beta_{j-2}} \: 
{a}^j (\ln Q^2; {\beta_2}, 
\ldots) \ .
\label{TPSbeta}
\ea
The MA couplings ${\A_1}(Q^2)$ and ${\tlA_1}(s)$
are finite in the IR (with the value $1/\beta_0$ at $Q^2=0$, 
or $s=0$) and show strong stability under the increase of the 
loop-level
$n_{\rm m} = j_{\rm max} - 1$ (see Figs.~\ref{shir1shir2}, \ref{shir3a}), 
and under the change of the 
renormalization scale (RScl) and scheme (RSch).
Another similar pQCD-approach is to analytize minimally
${\beta}({a})/{a} =
\partial \ln {a}(Q^2)/\partial \ln Q^2$
\cite{Neste2000,Neste2001,Neste2003}. This leads to an IR-divergent
analytic (${{\overline {\rm MA}}}$) coupling,
${\A_1}(Q^2) \sim (\Lambda^2/Q^2) (\ln (\Lambda^2/Q^2))^{-1}$
when $Q^2 \to 0$. At one-loop:
\ba
{\A_1}(Q^2) & = &
\frac{1}{\beta_0} \frac{ (Q^2/\Lambda^2) - 1}
{(Q^2/\Lambda^2) \ln (Q^2/\Lambda^2) } \ .
\ea
Also this coupling has improved stability under the
loop-level change, and under the RScl and RSch changes
(see Figs.~\ref{Nest20002001}, \ref{Nest2003}).
Numerical predictions of this model, at the 
one-loop level, for various observables, were
performed in Ref.~\cite{Neste2003}, and they
agree with the experimental results within the
experimental uncertainties and the theoretical uncertainties 
of the one-loop approximation. 


\section{Beyond the MA}
\label{sec:models}

The idea to make the QCD coupling 
{IR finite} phenomenologically is an old one,
by the substitution $\ln(Q^2/{\Lambda}^2) \mapsto
\ln[(Q^2 + 4 {m_g}^2)/{\Lambda}^2]$ where
$m_g$ is an effective gluon mass,
cf.~Refs.~\cite{Parisi:1980jy,Cornwall:1981zr,Mattingly:1993ej}.

On the other hand, the analytic MA, or
${\overline {\rm MA}}$, couplings 
can be {modified at low energies}, bringing in 
{additional parameter(s)} such that there is a possibility to
reproduce better a wide set of low energy QCD 
{experimental} data.

Among the recent proposed analytic couplings are:

1. Synthetic coupling proposed by Alekseev \cite{Alekseev:2005he}:
\begin{equation}
{\alpha}_{\mathrm{syn}}(Q^2)= {\alpha}^{\rm (MA)}(Q^2)
+\frac{\pi}{\beta_0}\left[ \frac{{c}{\Lambda}^2}{Q^2}-
\frac{{d} \; {\Lambda}^2}
{Q^2+{m_g}^2}\right], 
\label{Alekseev}
\end{equation}
where the three new parameters ${c}$, ${d}$ and
gluon mass ${m_g}$ were determined by requiring
${\alpha}_{\mathrm{syn}}(Q^2) - {\alpha}_{\mathrm{pt}}(Q^2) \sim
({\Lambda}^2/Q^2)^3$ (for the convergence of the gluon condensate)
and by the string condition ${V}(r) \sim {\sigma} r$ 
($r \to \infty$)
with ${\sigma} \approx 0.42^2 {\rm GeV}^2$.
This coupling is IR-divergent.

2. The coupling by Sriwastawa {\it et al.\/} \cite{Srivastava:2001ts}:
\ba
\lefteqn{
\frac{1}{{\alpha}^{(1)}_{\rm {SPPW}}(Q^2)} =
\frac{1}{{\alpha}^{(1)}_{\rm {SPPW}}({\Lambda}^2)} 
}
\nonumber\\
&&+ \frac{{\beta_0}}{\pi} \int\limits_{0}^{\infty}
\frac{(z-1)\, z^{p}}{(\sigma+z-i\varepsilon)(\sigma+1)
(1+z^{p})}
\,d \sigma ,
\label{Sriwas}
\ea
where $z=Q^2/{\Lambda^2}$ and 
$0 < p \leq 1$. This formula coincides
with Nesterenko's (one-loop) ${{\overline {\rm MA}}}$
coupling when~${p}=1$.

3. An IR-finite 
coupling 
proposed 
by Webber 
\cite{Webber:1998um}:
\begin{equation}
{\alpha}^{(1)}_{{W}}(Q^2) = \frac{\pi}{{\beta_0}} \left[
\frac{1}{\ln z} + \frac{1}{1-z}\,\frac{z+{b}}{1+{b}}
\left(\frac{1+{c}}{z+{c}}\right)^{\!{p}}\,
\right], 
\label{Webber}
\end{equation}
where $z=Q^2/\Lambda^2$ and specific values are chosen for 
parameters $b=1/4$, $c=4$, and $p=4$;
${\alpha}^{(1)}_{{W}}(0) \simeq \pi /(2 {\beta_0})$. 

4. ``Massive'' ${{\overline {\rm MA}}}$ or 
MA 
couplings ${\A_1}(Q^2)$ 
and ${\tlA_1}(s)$ proposed by
Nesterenko and Papavassiliou \cite{Neste2005}:
\ba 
{\tlA_1}^{\rm ({m})}(s) & = 
& \Theta(s - 4 {m}^2) {\tlA_1}(s) \ ,
\nonumber\\
{\A_1}^{\rm ({m})}(Q^2) 
& = &  
\frac{Q^2}{Q^2 + 4 {m}^2} \int_{4 {m}^2}^{\infty} 
{\rho_1}(\sigma) \frac{\sigma - 4 {m}^2}{\sigma + Q^2} 
\frac{d \sigma}{\sigma} \ ,
\nonumber\\
\label{massiveMA}
\ea
where ${m} \sim {{\overline \Lambda}}$; and
${\rho_1}(\sigma) = 
{\rho_1}^{\rm {(pt)}}(\sigma)$
in the MA case. In this case: 
${\A_1}^{\rm ({m})}(0) =
{\tlA_1}^{\rm ({m})}(0) = 0$.
The mass ${m}$ is some kind of threshold,
and can be expected to be $\sim {m_{\pi}}$.  

5. Two specific models of IR-finite analytic coupling
\cite{Cvetic:2006mk,Cvetic:2006gc}:
on the time-like axis $s \equiv -Q^2 > 0$, 
the parturbative discontinuity function ${{\rho}_1}(s)$,
or equivalently ${{\tlA}_1}^{\rm {(MA)}}(s)$, 
was modified in the
in the IR regime
($s \sim \bL^2$). A first possibility (model {'M1'}):
\bap
{\tlA_1^{\rm {(M1)}}}(s) & = & 
{c_f \bM_r^2} \delta(s - {\bM_r^2})
\nonumber \\
&&+ {k_0} \Theta({\bM_0^2} - s) + 
\Theta(s - {\bM_0^2}) {\tlA_1^{\rm {(MA)}}}(s) \ ,
\eap
where ${c_f}$, ${k_0}$, 
${c_r= \bM_r^2/{\bL}^2}$,
${c_0 = \bM_0^2/{\bL}^2}$ are four dimensionless
parameters of the model, all $\sim 1$. 
One of them (${k_0}$) can be eliminated
by requiring the (approximate) merging of M1 with MA at
large $Q^2$: 
\begin{displaymath}
|{\A_1}^{\rm {(M1)}}(Q^2) - 
{\A_1}^{\rm {(MA)}}(Q^2)|
\sim ({\bL^2}/Q^2)^2. 
\end{displaymath}
The Euclidean ${\A_1^{\rm {(M1)}}(Q^2)}$ is
\ba
{\A_1^{\rm {(M1)}}}(Q^2) & = & 
{\A_1^{\rm {(MA)}}}(Q^2) + \Delta {\mathcal{A}_1^{\rm {(M1)}}}(Q^2) \ ,
\nonumber\\
{\Delta \mathcal{A}_1^{\rm {(M1)}}}(Q^2) & = & 
- \frac{1}{\pi} \int_{\sigma= 0}^{{\bM_0^2}}
\frac{d \sigma {\rho^{\rm {(pt)}}_1}(\sigma) }{(\sigma + Q^2)} +
{c_f} \frac{ {\bM_r^2} Q^2 }
{ \left( Q^2 + {\bM_r^2} \right)^2 } 
\nonumber\\
&& - {d_f} \frac{ {\bM_0^2} }
{ \left( Q^2 + {\bM_0^2} \right) } \ ,
\label{M1b}
\ea
where the constant ${d_f}$ is
\bep
{d_f} \equiv - {k_0} + 
\frac{1}{\pi} \int_{{\bM_0^2}}^{\infty} 
\frac{d \sigma}{\sigma} {\rho^{\rm {(pt)}}_1}(\sigma) \ .
\eep
Another, simpler, possibility is (model {'M2'}):
\ba
{\tlA_1^{\rm {(M1)}}}(s) & = & {\tlA_1^{\rm {(MA)}}}(s) + 
{c_v} \Theta({\bM_p^2} - s) \ ,
\\
{\A_1^{\rm {(M1)}}}(Q^2) & = & {\A_1^{\rm {(MA)}}}(Q^2) + 
{c_v}  \frac{ \bM_p^2 }{(Q^2 + {\bM_p^2})} \ ,
\label{M2}
\ea
where ${c_v}$ and 
${c_p = \bM_p^2/{\bL^2}}$ 
are the model parameters.

6. Those anQCD models which respect the OPE-ITEP condition
are presented in Sec.~\ref{sec:itep}.

\section{Analytization of {higher powers} 
${a^k} \mapsto {\A_k}$}
\label{sec:hipow}

In MA model, the construction is \cite{ShS,Milton:1997mi,Milton:2000fi,Sh}
(MSSSh: Milton, Solovtsov, Solovtsova, Shirkov):
\begin{equation}
{a^k}(Q^2) \: \mapsto \:
{\mathcal{A}^{\rm {(MA)}}_k}(Q^2) = 
\frac{1}{\pi}\int_0^\infty \frac{d\sigma}{\sigma+Q^2}\: 
{\rho_k^{\rm {(pt)}}}(\sigma) \ ,
\label{MAAkdisp}
\end{equation}
where $k=1,2,\ldots$; ${\rho_k^{\rm {(pt)}}}(\sigma)=
\text{Im}[{a^k}(-\sigma-i\epsilon)]$;
and ${a}$ is given, e.g., by Eq.~(\ref{apt}).
In other words, ``minimal analytization'' (MA)
is applied to each power $a^k$.

As a consequence, in MA we have \cite{Magradze}
\bap
\frac{\partial {\A^{\rm {(MA)}}_1}(\mu^2)}{\partial \ln \mu^2} & = &
-\!{\beta_0} {\A^{\rm {(MA)}}_2}(\mu^2)\!-\!{\beta_1} {\A^{\rm {(MA)}}_3}(\mu^2)\!-\!\cdots,
\nonumber\\
\frac{\partial^2 {\A^{\rm {(MA)}}_1} (\mu^2)}
{\partial (\ln \mu^2)^2} & = &
2 \beta_0^2 \A^{\rm (MA)}_3\!+\!5 \beta_0 \beta_1 \A^{\rm (MA)}_4\!+\!\cdots,
\eap
etc. This is so because ${a^k}$, and
consequently ${\rho^{\rm {(pt)}}_k}(\sigma)$, fulfill analogous 
{RGE}'s.

The approach (\ref{MAAkdisp}) of constructing $\A_k$'s ($k \geq 2$)
can be applied to a specific model only (MA).
In other anQCD models (i.e., for other $\A_1(Q^2)$), 
the discontinuity functions
$\rho_k$ ($k \geq 2$) are not known. We present an approach 
\cite{Cvetic:2006mk,Cvetic:2006gc} that is applicable to any anQCD model, 
and reduces to the above approach in the MA model.
We proposed to maintain the {scale (RScl)}
evolution of these (truncated) relations for
any version of anQCD 
\ba
\frac{\partial {\A_1}(\mu^2; {\beta_2}, \ldots)}{\partial \ln \mu^2} 
& = &
- \beta_0 \A_2 - \cdots - 
\beta_{n_{\rm m}-2} \A_{n_{\rm m}} \ ,
\nonumber\\
\frac{\partial^2 {\A_1}(\mu^2; {\beta_2}, \ldots)}
{\partial (\ln \mu^2)^2 } 
& = &
 2 \beta_0^2 \A_3\!+\!5 \beta_0 \beta_1 \A_4\!+\!\cdots 
+ \kappa^{(2)}_{n_{\rm m}} \A_{n_{\rm m}},
\nonumber\\
\label{anRGE}
\ea
etc. Eqs.~(\ref{anRGE}) define the couplings
$\A_k(Q^2)$ ($k \geq 2$). 
Further, the evolution under the {scheme (RSch)} changes
will also be maintained as in
the MA case (and in pQCD):
\ba
\frac{\partial {\A_1}(\mu^2; {\beta_2}, \ldots)}
{\partial {\beta_2}} & \approx &
\frac{1}{{\beta_0}} {\A_3} + 
\frac{{\beta_2}}{3 \beta_0^2} {\A_5} + \cdots
+ k^{(2)}_{n_{\rm m}} {\A_{n_{\rm m}}} ,
\nonumber\\
\label{anRGEsch}
\ea
analogously for $\partial {\A_1}/\partial {\beta_3}$,
etc. In our approach, the basic space-like quantities are
${\A_1}(\mu^2)$ 
of a given anQCD model (e.g., MA, M1, M2)
and its logarithmic derivatives
\be
{\tA_{n}}(\mu^2) \equiv 
\frac{ (-1)^{n-1}}{\beta_0^{n-1} (n-1)!} 
\frac{ \partial^{n-1} {{\A_1}}(\mu^2)}
{\partial (\ln \mu^2)^{n-1}} \ ,
\quad (n=1,2,\ldots),
\label{tAn}
\ee
whose pQCD analogs are
\be
{{\widetilde a}_n}(\mu^2) \equiv
\frac{ (-1)^{n-1}}{\beta_0^{n-1} (n-1)!} 
\frac{ \partial^{n-1} {a}(\mu^2)}{\partial (\ln \mu^2)^{n-1}} \ ,
\quad (n=1,2,\ldots).
\label{tan}
\ee

At loop-level three ($n_{\rm m}=3$),
where we include in RGE (\ref{pRGE}) term
with $j_{\rm max}=4$ (thus ${\beta_2}$), 
relations (\ref{anRGE}) are 
\begin{equation}
{{\tA_2}}(\mu^2) 
= {{\A_2}}(\mu^2) + \frac{{\beta_1}}{{\beta_0}}{{\A_3}}(\mu^2), 
\quad
{{\tA_3}}(\mu^2) 
= {{\A_3}}(\mu^2),
\label{tA2tA3}
\end{equation}
implying
\begin{equation}
{{\A_2}}(\mu^2) 
= {{\tA_2}}(\mu^2) - \frac{{\beta_1}}{{\beta_0}} {{\tA_3}}(\mu^2), 
\quad 
{{\A_3}}(\mu^2) 
= {{\tA_3}}(\mu^2).
\label{A2A3}
\end{equation}
The RSch (${\beta_2}$) dependence is obtained from
the truncated Eqs.~(\ref{anRGEsch}) and (\ref{anRGE})
\be
\frac{\partial {\tA_j}(\mu^2;{\beta_2})}
{\partial {\beta_2}} \approx
\frac{1}{2 \beta_0^3} 
\frac{\partial^2 {\tA_j}(\mu^2; {\beta_2})}
{\partial (\ln \mu^2)^2} \ ,
\label{dA1dbll3}
\ee
where $(j=1,2,\ldots)$ and ${\tA_1} \equiv {\A_1}$.

At loop-level four (${n_{\rm m}=4}$), 
where we include in {RGE} (\ref{pRGE}) term
with ${j_{\rm max}=5}$ 
(thus ${\beta_3}$), relations analogous to
(\ref{A2A3})-(\ref{dA1dbll3}) can be found
\cite{Cvetic:2006gc}. 

\begin{figure}[htb] 
\centering
\epsfig{file=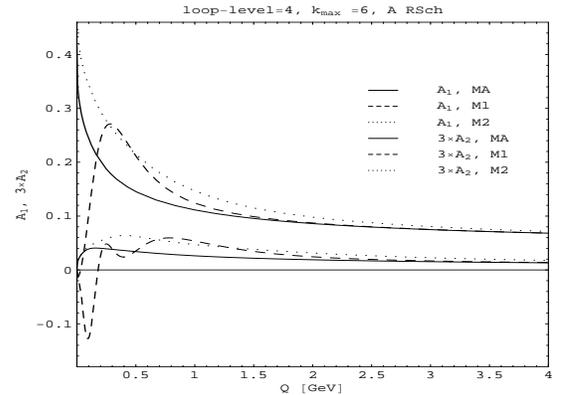,width=8.2cm,height=6.cm}
\vspace{-0.6cm}
\caption{\footnotesize
${\A_1}$ and ${\A_2}$ for various models 
(M1, M2 and MA) 
with specific model parameters: ${c_0}=2.94$, 
${c_r}=0.45$, 
${c_f}=1.08$ for M1;
${c_v}=0.1$, ${c_p}=3.4$ for M2; 
${n_f}=3$, ${\bL_{(n_f=3)}}=0.4$ GeV
in all three models. The upper three curves are ${\A_1}$,
the lower three are $3 \times {\A_2}$. All couplings  
are in v-scheme (see Subsec.~\ref{subsec:lbe}). 
${\A_2}$ is constructed with our approach.
Figure from: Ref.~\cite{Cvetic:2006gc}.}
\label{M1M2Ma}
\end{figure}

It turns out that there is a clear hierarchy in
magnitudes $|\A_1(Q^2)| > |\A_2(Q^2)| > |\A_3(Q^2)| > \cdots$
at all $Q^2$, in all or most of the anQCD models
(cf.~Fig.~\ref{M1M2Ma} for MA, M1, M2; and 
Fig.~\ref{ourmodel} in Sec.~\ref{sec:itep} for another model).

We recall that the perturbation series of a space-like
observable ${\cal D}(Q^2)$ ($Q^2 \equiv - q^2 > 0$)
can be written as
\ba
{{\cal D}}(Q^2)_{\rm pt} &=& a + 
d_1 a^2 +  d_2 a^3 + \cdots ,
\label{Dpt1a}
\\
&=& {\widetilde a}_1 + d_1 {\widetilde a}_2 + 
\left( d_2 - \frac{\beta_1}{\beta_0} d_1 \right) 
{\widetilde a}_3  + \cdots ,
\label{Dpt1b}
\ea
where the second form (\ref{Dpt1b})
is the reorganization of the
perturbative power expansion (\ref{Dpt1a})
into a perturbation expansion in terms of 
${\widetilde a}_n$'s
(\ref{tan}) (note: ${\widetilde a}_1 \equiv a$).
The basic analytization rule we adopt is the
replacement
\be
{{\widetilde a}_n} \mapsto {\tA_n}
\qquad (n=1,2,\ldots) \ ,
\label{basican}
\ee
term-by-term in expansion (\ref{Dpt1b}),
and this is equivalent to the analytization rule 
$a^n \mapsto \A_n$ term-by-term in expansion (\ref{Dpt1a}).
However, in principle, 
other analytization procedures could be adopted,
e.g. $a^n \mapsto \A_1^n$, or $a^n \mapsto
\A_1 \A_{n-1}$, etc.
The described analytization $a^n \mapsto \A_n$ reduces to the
MSSSh analytization in the case of the MA model (i.e., in
the case of $\A_1 = \A_1^{({\rm MA})}$), because the
aforementioned RGE-type relations hold also in the MA case.

Let's denote by ${\cal D}^{(n_{\rm m})}(Q^2)$
the TPS of (\ref{Dpt1a})
with terms up to (and including) the term
$\sim a^{n_{\rm m}}$, and by ${\cal D}_{\rm an.}^{(n_{\rm m})}(Q^2)$
the corresponding truncated analytic series (TAS) obtained from
the previous one by the term-by-term analytization $a^n \mapsto \A_n$.
The evolution of $\A_k(Q^2)$ under the 
changes of the RSch was truncated in such a way that
$\partial {{\cal D}}_{\rm an.}^{({n_{\rm m}})}(Q^2)/\partial {\beta_j}
\sim {\A}_{{n_{\rm m}}+1}$ (where $j \geq 2$).
Further, our definition of
${\A_k}$'s (${k} \geq 2$) via Eqs.~(\ref{anRGE})
[cf.~Eqs.~(\ref{A2A3})] 
involves truncated series which, however, still
ensure the ``correct'' RScl-dependence
$\partial {{\cal D}}_{\rm an.}^{({n_{\rm m}})}(Q^2)/\partial {\mu^2}
\sim {\A}_{{n_{\rm m}}+1}$.
This is all in close analogy with the pQCD results for TPS's:
$\partial {{\cal D}}^{({n_{\rm m}})}(Q^2)/\partial {\beta_j}
\sim a^{{n_{\rm m}}+1}$, and
$\partial {{\cal D}}^{({n_{\rm m}})}(Q^2)/\partial \mu^2
\sim a^{{n_{\rm m}}+1}$. In conjunction with the mentioned
hierarchy depicted in Fig.~\ref{M1M2Ma}, this means that 
the evaluated TAS will have increasingly weaker RSch and RScl dependence 
when the number of TAS terms increases,
at all values of $Q^2$. 

On the other hand, if the analytization of powers
were performed by another rule, for example,
by the simple rule $a^n \mapsto \A_1^n$, the above
RScl$\&$RSch-dependence of the TAS would not be valid any more.
An increasingly weaker RScl$\&$RSch-dependence of TAS (when the number of
TAS terms is increased) would not be guaranteed any more.

\section{Calculation of ${\A_{\nu}}$ for
${\nu}$ noninteger}
\label{sec:nonint}

Analytization of noninteger powers in MA model was performed and
used in Refs.~\cite{Bakulev}, representing a generalization 
of results of Ref.~\cite{Broadhurst:2000yc}. The approach was
motivated by a previous work \cite{KS2001} where MA-type of
analytization of expressions for hadronic observables was postulated,
these being integrals linear in $a(t Q^2)$ [similar to the
dressed gluon approximation expressions, cf.~Eq.~(\ref{LS1})
and the first line of Eq.~(\ref{TASbv})].
Analytization of noninteger powers $a^{\nu}$ or $a^{\nu} \ln a$, 
is needed in calculations of pion electromagnetic form factor, 
and in some resummed expressions for Green functions or observables, 
calculated within an anQCD model.

In the mentioned approach, use is made of the Laplace transformation
$(f)_{L}$ of function $f$
\bep
f(z) \mapsto (f)_{L}(t): \quad f(z)=\int_0^{\infty} dt e^{-z t} (f)_{L}(t)
\ ,
\eep
where $z \equiv \ln(Q^2/\Lambda^2)$.  
Using notations (\ref{tan}) and (\ref{tAn}),
it can be shown
\ba
( {{\widetilde a}_n})_{L}(t) & = & 
\frac{t^{n-1}}{\beta_0^{n-1} (n-1)!} 
({a})_{L}(t) \ ,
\label{tanL}
\\
( {{\widetilde \A}_n})_{L}(t) & = & 
\frac{t^{n-1}}{\beta_0^{n-1} (n-1)!} 
({\A_1})_{L}(t) \ .
\label{tAnL}
\ea
Therefore, it is natural to define for any real ${\nu}$
the following Laplace transforms:
\ba
( {{\widetilde a}_{\nu}})_{L}(t) & = & 
\frac{t^{{\nu}-1}}{\beta_0^{{\nu}-1} 
\Gamma({\nu}) } ({a})_{L}(t) \ ;
\label{tanuL}
\\
( {{\widetilde \A}_{\nu}})_{L}(t) & = & 
\frac{t^{{\nu}-1}}{\beta_0^{{\nu}-1} 
\Gamma({\nu}) } ({\A_1})_{L}(t) \ .
\label{tAnuL}
\ea
In MA model, at one-loop level, $({a})_{L}(t)$
and $({\A_1})_{L}(t)$ are known
\ba
{a}(z) & = & \frac{1}{{\beta_0} z}
\ \Rightarrow \ ({a})_{L}(t) = \frac{1}{{\beta_0}} \ .
\label{aLt}
\\
{\A_1}(z) & = & \frac{1}{{\beta_0}}\left(
\frac{1}{z} - \frac{1}{e^z - 1} \right) \ \Rightarrow
\nonumber\\
({\A_1})_{L}(t) & = & \frac{1}{{\beta_0}}
\left( 1 - \sum_{k=1}^{\infty} \delta(t - k) \right) \ .
\label{A1Lt}
\ea
Since at one-loop ${\widetilde \A}_{\nu} = {\A}_{\nu}$,
it follows in one-loop MA model
\be
{\A}_{\nu}(z) 
=
\int_0^{\infty} dt e^{-z t} \frac{t^{{\nu}-1}}
{\beta_0^{\nu} \Gamma({\nu}) }
\left( 1\!-\!\sum_{k=1}^{\infty} \delta(t - k) \right).
\label{BNS}
\ee
Similarly, since
\bep
{a^{\nu}}(z) \ln {a}(z) = 
\frac{d}{d {\nu}}{a^{\nu}}(z) \ ,
\eep
it can be defined
\be
\left[ \frac{d}{d {\nu}} {a^{\nu}}(z) 
\right]_{\rm MA} \equiv 
\frac{d}{d {\nu}} {\A_{\nu}}(z) \ .
\label{MAanulna}
\ee
To calculate higher (two-)loop level $\A_{\nu}(z)$
in MA model, the authors of Refs.~\cite{Bakulev}
expressed the two-loop
${a_{(2)}}(z)$ in terms of one-loop powers
${a_{(1)}}^m(z) \ln^n {a_{(1)}}(z)$
and then followed the above procedure.

\section{Evaluation methods for observables}
\label{sec:eval}

In pQCD, the most frequent method of evaluation of the
leading-twist part of a space-like physical quantity is the 
evaluation of the available (RG-improved)
truncated perturbation
series (TPS) in powers of perturbative coupling $a$.
Within the anQCD models, an analogous method
is the aforementioned replacement $a^n \mapsto \A_n$ in the
TPS (where $\A_n$ are constructed in Sec.~\ref{sec:hipow}), 
and the evaluation thereof.
More specifically, consider an observable ${{\cal D}}(Q^2)$ 
depending on a single space-like physical scale $Q^2 (\equiv - q^2) >0$.
Its usual perturbation series has the form (\ref{Dpt1a}),
where ${a} = {a}(\mu^2;{\beta_2}, {\beta_3},\ldots)$,
with $\mu^2 \sim Q^2$. 
For each TPS ${{\cal D}}(Q^2)_{\rm pt}^{({N})}$
of order $N$, in the minimal anQCD (MA) model, 
the authors MSSSh
\cite{Milton:1997mi,Milton:2000fi,Sh}
introduced the aforementioned replacement
$a^n \mapsto \A_n^{({\rm MA})}$:
\be
{{\cal D}}(Q^2)_{\rm an}^{({N}) \rm (MSSSh)} =
{\A_1}^{({\rm MA})} + {d_1} {\A_2}^{({\rm MA})} +
\cdots {d_{N-1}} {\A_N}^{({\rm MA})} \ .
\label{MSSShTPS}
\end{equation}
This method of evaluation (via $a^n \to {\A}_n$)
was extended to any anQCD model in
\cite{Cvetic:2006mk,Cvetic:2006gc}
(cf.~Sec.~\ref{sec:hipow}). Further,
in the case of inclusive space-like observables,
the evaluation was extended 
to the resummation of the large-${\beta_0}$ terms:

\subsection{Large-${\beta_0}$-motivated expansion of observables}
\label{subsec:lbe}

We summarize the presentation of Ref.~\cite{Cvetic:2006gc}.
We work in the RSch's where each ${\beta_k}$ 
(${k} \geq 2$)
is a polynomial in ${n_f}$ of order ${k}$; 
in other words, it is
a polynomial in ${\beta_0}$:
\be
{\beta_k} = 
\sum_{j=0}^{k} {b_{kj}} \beta_0^j \ , 
\qquad {k}=2,3,\ldots
\label{betak}
\ee
The ${\bMS}$ belongs to this class of schemes. 
In such schemes, the coefficients ${d_n}$ 
of expansion (\ref{Dpt1a}) 
have the following specific form in terms of ${\beta_0}$:
\ba
\lefteqn{
{{\cal D}}(Q^2)_{\rm pt} = 
a + (c_{11} {\beta_0} + c_{10}) a^2}  
\nonumber\\
&&+(c_{22} \beta_0^2 + c_{21} \beta_0 + c_{20} + c_{2,-1} \beta_0^{-1}) a^3 
+ \cdots .
\label{Dpt2}
\ea
We can construct a separation of this series into a sum of two
{RScl}-independent terms --  
the leading-${\beta_0}$ (${\rm L}{\beta_0}$),
and beyond-the-leading-${\beta_0}$ (${\rm BL}{\beta_0}$)
\ba
{{\cal D}}_{\rm pt} &=& 
{{\cal D}}^{({\rm L}{\beta_0})}_{\rm pt} + {{\cal D}}^{({\rm BL}{\beta_0})}_{\rm pt} \ ,
\label{lb0nlb0}
\ea
where 
\ba
\lefteqn{
{{\cal D}}^{({\rm L}{\beta_0})}_{\rm pt} = 
{a} + {a}^2 \left[ \beta_0 c_{11} \right] + 
{a}^3 \left[ \beta_0^2 c_{22}\!+\!\beta_1 c_{11} \right]} 
\nonumber\\
&& +{a}^4 \left[ \beta_0^3 c_{33}\!+\!\frac{5}{2} {\beta_0} {\beta_1} 
c_{22}\!+\!\beta_2 c_{11} \right] + {\cal {O}}(\beta_0^4 {a}^5).
\label{D1}
\ea 
Expression (\ref{D1}) is not the standard leading-$\beta_0$ 
contribution, since it contains also
terms with $\beta_j$ ($j \geq 1$), but only in
a minimal way to ensure that the expression
contains all the leading-$\beta_0$ terms and
at the same time remains RScl-independent.
It can be shown that, 
for inclusive observables, all the coefficients in
this ${\rm L}{\beta_0}$ contribution can be obtained,
and can be expressed in the integral form \cite{Neubert}
\be
{{\cal D}}^{({\rm L}{\beta_0})}(Q^2)_{\rm pt} = 
\int_0^\infty \frac{dt}{t}\: 
{F}_{{\cal D}}^{\cal {E}}(t) \: 
{a}(t e^{{\cal C}} Q^2) \ ,
\label{LS1}
\ee
where ${F}_{{\cal D}}^{\cal {E}}(t)$ is the (Euclidean) ${\rm L}{\beta_0}$ -characteristic
function. In ${\bMS}$ scheme, 
${\Lambda} = {\bL}$ which corresponds here to 
${{\cal C}}={{\overline {\cal C}}} \equiv -5/3$. 
No RScl ${\mu^2}$ appears in (\ref{LS1}).
Expression (\ref{LS1}) is referred to in the literature sometimes
as dressed gluon approximation.

The ${\rm BL}{\beta_0}$ contribution is usually known 
only to $\sim {a}^3$ or
$\sim {a}^4$. For it, we can use an arbitrary RScl 
${\mu^2} \equiv Q^2 e^{C} \sim Q^2$. 
Further, the powers ${a}^k$ can be
reexpressed in terms of ${{\widetilde a}_n}(\mu^2)$ (\ref{tan}):
\be
{a}^2 = {{\widetilde a}_2}\!-\!({\beta_1}/{\beta_0})
{{\widetilde a}_3}\!+\!\cdots ,
\quad
{a}^3 =  {{\widetilde a}_3}\!+\!\cdots .
\label{ata}
\ee
Therefore,
\ba
\lefteqn{
{{\cal D}}(Q^2)_{\rm (TPS)} = 
{{\cal D}}^{({\rm L}{\beta_0})}(Q^2)_{\rm pt}}  
\nonumber\\
&&+{\tlt_2} \; 
{ {\widetilde a}_2}(Q^2 e^{C}) + 
{\tlt_3} \;
{{\widetilde a}_3}(Q^2 e^{C}) +
{\tlt_4}  \;
{{\widetilde a}_4}(Q^2 e^{C}), 
\label{TASbvgen}
\ea
where ${\tlt_2}=c_{10}$ is scheme-independent, 
and coefficients ${\tlt_3}$ and ${\tlt_4}$ 
have a scheme dependence (depend on 
${\beta_2}$, ${\beta_3}$ -- i.e., 
on ${b_{2j}}$ and ${b_{3j}}$).
We note that expression (\ref{TASbvgen}) is
not really a pure TPS, because its ${\rm L}{\beta_0}$
contribution (\ref{D1}) is not truncated.
An observable-dependent scheme (D-scheme) 
can be chosen such that ${\tlt_3}={\tlt_4=0}$.
For the Adler function ${\cal D}=d_v$, such a scheme will
be called v-scheme.
The analytization of the obtained ${{\cal D}}(Q^2)_{\rm (TPS)}$
(\ref{TASbvgen})
is performed by the substitution 
${{\widetilde a}_n} \mapsto {{\tA_n}}$,
Eq.~(\ref{basican}), leading to the
truncated analytic series (TAS)
\ba
\lefteqn{
{{\cal D}}(Q^2) =  
{{\cal D}}(Q^2)_{{\rm (TAS)}} + 
{\cal {O}}(\beta_0^3 \tA_5) 
\ ,}
\label{TASav}
\\
\lefteqn{
{{\cal D}}(Q^2)_{{\rm (TAS)}} = 
\int_0^\infty \frac{dt}{t}\: 
{F}_{{\cal D}}^{\cal {E}}(t) \: 
{\A_1} (t e^{{\cal C}} Q^2)
}
\nonumber\\
&& +\!c_{10} {\tA_2}(Q^2 e^{{C}})\!+\!{\tlt_3}{\tA_3}
(Q^2 e^{C})\!+\!{\tlt_4} {\tA_4}(Q^2 e^{C}). 
\label{TASbv}
\ea
In the D-scheme, the last two terms disappear.
Eq.~(\ref{TASbv}) is a method that one can use to
evaluate any inclusive space-like QCD observable in any
anQCD model. 
As argued in Sec.~\ref{sec:hipow}, 
the scale and scheme dependence of the TAS is
very suppressed
\be
\frac{\partial {{\cal D}}(Q^2)_{\rm (TAS)}}{\partial X}
\sim \beta_0^3 {\tA_{5}} \sim 
\beta_0^3 {\A_{5}} \quad  
(X = \ln {\mu^2}, {\beta_j}) \ .
\label{RSchdep}
\ee
If the BL$\beta_0$ perturbative contribution is known
exactly only up to (and including) $\sim\!a^3$, then
no ${\tlt_4}$ term appears in Eq.~(\ref{TASbv}) and
the precision in Eqs.~(\ref{TASav}) and (\ref{RSchdep}) is diminished:
${\cal O}( \beta_0^3 {\A_{5}}) \mapsto 
{\cal O}( \beta_0^2 {\A_{4}})$.

It is interesting to note that the Taylor expansion
of ${\A_1} (t e^{{\cal C}} Q^2)$ in
${{\cal D}}^{({\rm L}{\beta_0})}(Q^2)_{\rm an}$ 
in (\ref{TASbv}) around a chosen RScl $\ln(\mu^2)$
reveals just the aforementioned $a^n \mapsto \A_n$ 
analytization of the
large-${\beta_0}$ part (\ref{D1}), in any anQCD:
\bap
{{\cal D}}^{({\rm L}{\beta_0})}_{\rm an} &=& 
\int_0^\infty \frac{dt}{t}\: 
{F}_{{\cal D}}^{\cal {E}}(t) \: 
{\A_1} (t e^{{\cal C}} Q^2)
\nonumber\\
&=&
{\A_1} + {\A_2} \left[ \beta_0 c_{11} \right] + 
{\A_3} \left[ \beta_0^2 c_{22}\!+\!\beta_1 c_{11} \right] 
\nonumber\\
&& +
{\A_4} \left[ \beta_0^3 c_{33}\!+\!\frac{5}{2} \beta_0 \beta_1 
c_{22}\!+\!\beta_2 c_{11} \right] + {\cal {O}}(\beta_0^4 {\A_5}),
\eap
where ${\A_k} = {\A_k}(\mu^2;{\beta_2},
{\beta_3},\ldots)$. In other words, at the leading-$\beta_0$
level, the natural analytization $a \mapsto \A_1$ in integral
(\ref{LS1}) is equivalent to the term-by-term analytization
$a^n \mapsto \A_n$ ($\Leftrightarrow {\widetilde a}_n \mapsto
{\widetilde \A}_n$) in the corresponding perturbation series. 
This thus represents yet another motivation for the
analytization $a^n \mapsto \A_n$ 
[$\Leftrightarrow$ Eq.~(\ref{basican}) postulated in Sec.~\ref{sec:hipow}]
of {\em all} the available perturbation terms in ${\cal D}$.
For the first motivation, based on the systematic weakening of the
RScl$\&$RSch dependence of the truncated analytized ${\cal D}$,
see the end of Sec.~\ref{sec:hipow}.   

\subsection{Applications in phenomenology}
\label{subsec:phen}

Evaluations in MA model, with the MSSSh-approach
$a^n \mapsto \A_n^{({\rm MA})}$
\cite{Sh,Milton:1997mi,Milton:2000fi}, 
are usually performed in ${\bMS}$ scheme.
The only free parameter
is ${ \Lambda}$ ($= { {\overline \Lambda}}$).
Fitting the experimental data for 
{$\Upsilon$-decay, $Z \to$ hadrons, $e^+e^- \to$ hadrons},
to the MSSSh approach for MA
at the two- or three-loop level, they obtained
${ \Lambda}_{n_f=5} \approx 0.26$-$0.30$ GeV,
corresponding to:  ${ \Lambda}_{n_f=3} \approx 0.40$-$0.44$
GeV, and $\pi {\A_1}^{{\rm (MA)}}(M_Z^2) \approx
0.124$, which is above the {pQCD} world-average
value ${\alpha_s}(M_Z^2) \approx 0.119 \pm 0.001$.
The apparent convergence of the MSSSh nonpower
truncated series is also remarkable -- see Table \ref{t1}.
\begin{table}
\caption{\label{t1}Various order contributions to observables
within PT, and MSSSh (=APT) methods \cite{Sh,Shirkov:2006gv}:}
\begin{ruledtabular}
\begin{tabular}{ccccc}\hline
Process&Method&1st order&2nd &3rd 
\cr
\hline
GLS &PT & 65.1\% & 24.4\% &10.5\%
\cr
($Q \sim 1.76$GeV) &APT & 75.7\% & 20.7\% &3.6\%
\cr
\hline
$r_{\tau}$ &PT & 54.7\% & 29.5\% &15.8\%
\cr
($M_{\tau}=1.78$GeV) &APT & 87.9\% & 11.0\% &1.1\%
\cr
\end{tabular}
\end{ruledtabular}
\end{table}

In Refs.~\cite{Cvetic:2006mk,Cvetic:2006gc}, the
aformentioned TAS evaluation method (\ref{TASbv})
in anQCD models MA (\ref{MAA1disp}), M1 (\ref{M1b})
and M2 (\ref{M2}) was applied to the inclusive observables
Bjorken polarized sum rule (BjPSR) ${d}_b(Q^2)$,
Adler function ${d}_v(Q^2)$ and semihadronic $\tau$
decay ratio $r_{\tau}$
The exact values of coefficients $d_1$ and $d_2$
are known for space-like observables BjPSR 
${d}_b(Q^2)$ \cite{LV} and
(massless) {Adler} function 
${d}_v(Q^2)$ \cite{d1,d2}. (The exact coefficient
$d_3$ of ${d}_v$ has been recently obtained \cite{d3}, but
was not included in the analysis of 
Ref.~\cite{Cvetic:2006gc} that we present here; rather,
an estimated value of $d_3$ was used.)
In the v-scheme, the evaluated massless ${d}_v(Q^2)$ is
\ba
{d}_v(Q^2)_{({\rm TAS})} &=&  
\int_0^\infty \frac{dt}{t}\: {F}_{v}^{\cal {E}}(t) \: 
{\A_1} (t e^{{\overline {\cal C}}} Q^2; 
{\beta_2}^{\rm (x)}, {\beta_3}^{\rm (x)}) 
\nonumber\\
&& + 
\frac{1}{12} {\tA_2}( e^{{\overline {\cal C}}} Q^2 ) \ ,
\label{dvARSch}
\ea
while BjPSR ${d}_b(Q^2)_{({\rm TAS})}$ has one more term
${\widetilde t}_3 {\tA_3}( e^{{\overline {\cal C}}} Q^2 )$.
The difference between the (massless) true
${d}_x(Q^2)$ ($x=v,b$) and 
${d}_x(Q^2)_{({\rm TAS})}$ 
is ${\cal O}(\beta_0^2 \tA_4)$.   
The semihadronic $\tau$ decay ratio $r_{\tau}$ is, 
on the other hand,
a time-like quantity, but can be expressed as a contour integral
involving the Adler function ${d}_v$: 
\ba
\lefteqn{
 {r}_{\tau}(\Delta S\!=\!0,m_q\!=\!0) =}
\nonumber\\
&& \frac{2}{\pi} \int_0^{{m_{\tau}}^2} \frac{d s}{{m_{\tau}}^2}
\left( 1 - \frac{s}{{m_{\tau}}^2} \right)^2
\left( 1 + 2 \frac{s}{{m_{\tau}}^2} \right) 
\rm{Im} { \Pi}(s) =
\nonumber\\ 
&&  \frac{1}{2 \pi}
\int_{-\pi}^{+\pi} d \phi \: (1 + e^{i \phi})^3 (1 -e^{i \phi})
{d}_v(Q^2={m_{\tau}}^2 e^{i \phi}) .
\label{rtaudv}
\ea
This implies for the leading-${\beta_0}$ term of 
${r}_{\tau}$
\be
{r}_{\tau}(\Delta S\!=\!0,m_q\!=\!0)^{\rm ({\rm L}{\beta_0})} = 
\int_0^\infty \frac{dt}{t}\: {F}_{r}^{\cal {M}}(t) \: 
{\tlA_1} (t e^{{\overline {\cal C}}} 
{m_{\tau}}^2) ,
\label{LSrt}
\ee
where ${\tlA_1}$ is the time-like coupling appearing
in Eqs.~(\ref{tlA1A1})-(\ref{tlA1rho1}), and superscript
${\cal {M}}$ in the characteristic function indicates that it is
Minkowskian (time-like). The latter was obtained by Neubert
(second entry of Refs.~\cite{Neubert}).
The beyond-the-leading-${\beta_0}$ (${\rm BL}{\beta_0}$) 
contribution is the contour integral 
\ba
\lefteqn{
{r}_{\tau}(\triangle S\!=\!0, m_q\!=\!0)^{({\rm BL}{\beta_0})} =}
\nonumber\\
&&\frac{1}{24 \pi}
\int_{-\pi}^{+\pi} d \phi \: (1\!+\!e^{i \phi})^3 (1\!-\!e^{i \phi})
{\tA_2}( e^{{\overline {\cal C}}} {m_{\tau}}^2 e^{i \phi}). 
\label{rtbLS}
\ea
\begin{table}
\caption{\label{t2}
Results of evaluation of
${r}_{\tau}(\triangle S = 0,m_q=0)$ and of 
{BjPSR} 
${d}_b(Q^2)$ ($Q^2=2$ and $1 {\rm GeV}^2$),
in various {anQCD} models,
using TAS method (\ref{TASbv}).
The experimental values are 
${r}_{\tau}(\triangle S = 0,m_q=0)
= 0.204 \pm 0.005$, 
${d}_b(Q^2=2 \ {\rm GeV}^2) = 0.16 \pm 0.11$
and ${d}_b(Q^2=1 \ {\rm GeV}^2) = 0.17 \pm 0.07$.}
\begin{ruledtabular}
\begin{tabular}{llll}
 & ${r}_{\tau}$ &
${d}_b(Q^2=2)$ &
${d}_b(Q^2=1)$ 
\\
\hline
MA & 0.141 & 0.137 & 0.155
\\
M1 & 
0.204 & 0.160 & 0.170
\\
M2 & 
0.204 & 0.189 & 0.219 
\end{tabular}
\end{ruledtabular}
\end{table}
\begin{figure}[htb] 
\centering
\epsfig{file=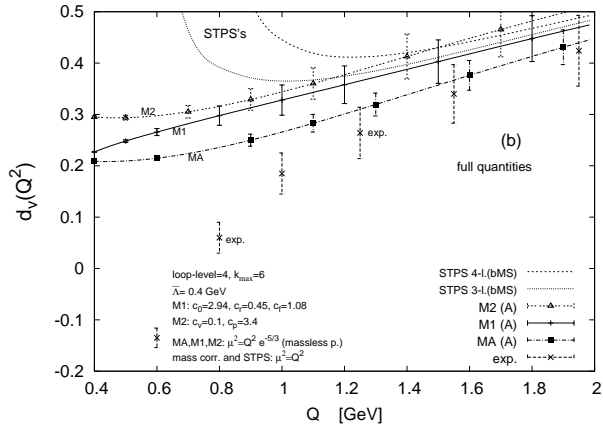,width=8.2cm}
\vspace{-0.3cm}
\caption{\footnotesize
Adler function as predicted by pQCD,
and by our approach in several anQCD models: {MA, M1, M2}.
The full quantity is depicted, 
with the contribution of massive quarks included.
The experimental values are from \cite{Eidelman}.
Figure from: Ref.~\cite{Cvetic:2006gc}.}
\label{Adler}
\end{figure}
The parameters of anQCD models M1 (\ref{M1b}) and M2 (\ref{M2}) 
were then determined \cite{Cvetic:2006gc}
by fitting the
evaluated observables to the experimental central values
${r}_{\tau}(\triangle S=0,m_q=0) = 0.204$ 
(for M1 and M2), and
to ${d}_b(Q^2=1 {\rm GeV}^2) = 0.17$ and 
${d}_b(Q^2=2) = 0.16$ (for M1).
For M1 we obtained:  ${c_f}=1.08$, 
${c_r}=0.45$, ${c_0}=2.94$.
For M2 we obtained:  
${c_v}=0.1$ and ${c_p}=3.4$.

\begin{figure}[htb] 
\centering
\epsfig{file=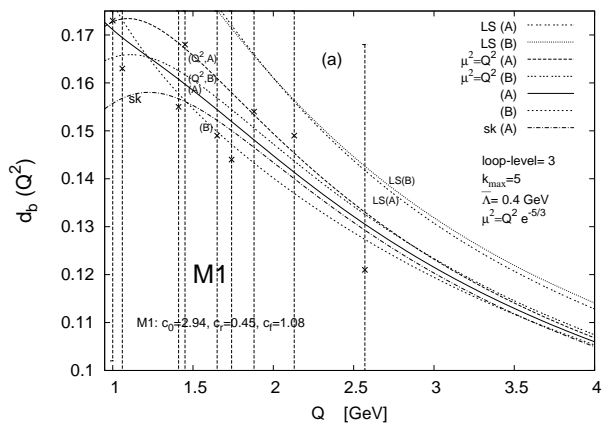,width=8.2cm}
\vspace{-0.3cm}
\caption{\footnotesize
Bjorken polarized sum rule ({BjPSR})
${d}_b(Q^2)$ in model M1, 
in various {RSch's} and at various {RScl's}. 
The vertical lines represent
experimental data, with errorbars in general covering the entire
depicted range of values.}
\label{BjPSRM1}
\end{figure}

\begin{figure}[htb] 
\centering
\epsfig{file=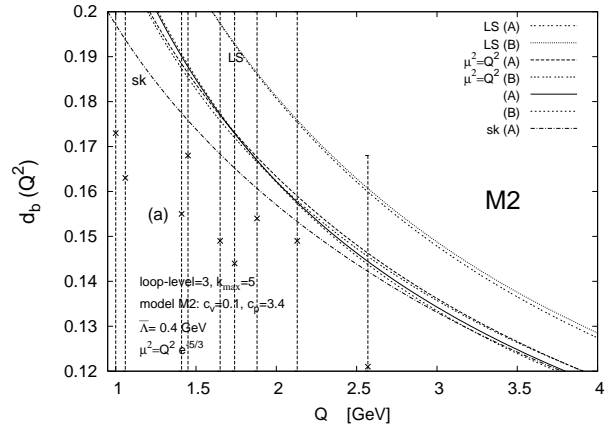,width=8.2cm}
\vspace{-0.3cm}
\caption{\footnotesize
As in the Fig.~\ref{BjPSRM1}, but this time for model M2.
Both figures from: Ref.~\cite{Cvetic:2006gc}.}
\label{BjPSRM2}
\end{figure}

The numerical results were then obtained \cite{Cvetic:2006gc}.
In models MA, M1 and M2 they are given for $r_{\tau}$ in Table \ref{t2},
for Adler function $d_v(Q^2)$ in Fig.~\ref{Adler},
and for BjPSR $d_b(Q^2)$ (in M1 and M2)
in Figs.~\ref{BjPSRM1} and \ref{BjPSRM2} (Table \ref{t2} and
Figs.~\ref{Adler}, \ref{BjPSRM1}, \ref{BjPSRM2} are taken
from Ref.~\cite{Cvetic:2006gc}). All results were calculated in
the v-scheme. For details, we refer to Ref.~\cite{Cvetic:2006gc}.

Analytic QCD models have been used also in
the physics of mesons \cite{mes1,mes2},
in calculating various meson masses by summing
two contributions: that of the confining part 
and that of the (one-loop) perturbative part of
the Bethe-Salpeter potential. In Refs.~\cite{mes1},
the (one-loop) MA coupling \cite{ShS} was used 
to calculate/predict the masses;
in Refs.~\cite{mes2}, the experimental mass spectrum
was used to extract the approximate values of the
(analytic) coupling $\A_1(Q^2)$ at low $Q^2$.
In this formalism, the current quark masses were replaced 
by the constituent quark masses, accounting in this
way approximately for the quark self-energy effects.
The results by the authors of Ref.~\cite{mes2}
indicate that $\A_1(Q^2)$ remains finite (and becomes
possibly zero) when $Q^2 \to 0$.

\section{Analytic QCD and {ITEP-OPE} philosophy}
\label{sec:itep}

In general, the deviations of analytic
${\A_1}(Q^2)$ from the perturbative coupling 
${a}_{\rm pt}(Q^2)$ at high $Q^2 \gg {\Lambda}^2$
are power terms
\be
|\delta {\A_1}(Q^2)| \equiv
|{\A_1}(Q^2) - {a}_{\rm pt}(Q^2)| \sim 
\left( \frac{{\Lambda}^2}{Q^2} \right)^{{k}}
\quad (Q^2\!\gg\!{\Lambda}^2),
\nonumber
\ee
where ${k}$ is a given positive integer. 
Such a coupling introduces in the evaluation (of the leading-twist)
of inclusive space-like observables ${ {\cal D}}(Q^2)$, already at the
leading-${\beta_0}$ level, an {UV}
contribution $\delta { {\cal D}}^{\rm ({UV})}(Q^2)$
which behaves like a power term \cite{Cvetic:2007ad}
\be
\delta { {\cal D}}^{\rm ({UV})}(Q^2) \sim 
\left( \frac{{ \Lambda}^2}{Q^2} \right)^{\rm min({k,n})} 
\quad {\rm if} \ {k \not= n} \ ,
\label{dDUV}
\ee
where ${n} \ \epsilon \ {\cal N}$ is the position of the leading
{IR} renormalon of the observable ${ {\cal D}}(Q^2)$;
if $k=n$, then the left-hand side of Eq.~(\ref{dDUV}) changes to
$( \Lambda^2/Q^2 )^n \ln ( \Lambda^2/Q^2 )$
\cite{Cvetic:2007ad}. 
Such nonperturbative contributions coming from the
{UV} sector contradict the 
{ITEP Operator Product Expansion (OPE)}
philosophy (the latter saying that such terms can come {only}
from the {IR} sector) \cite{DMW}.

Two specific sets of models of anQCD have been introduced 
in the literature so far
such that they do not contradict the {ITEP-OPE}: 

(A) a model set based on a modification of the ${\beta}({a})$ function 
\cite{Raczka}; 

(B) a model set obtained by a direct construction
\cite{Cvetic:2007ad}. 

\subsection{Set of models A}
\label{subsec:Raczka}

This is the set of models constructed in Refs.~\cite{Raczka}.
The {TPS} ${\beta}({a})$
used in pQCD is
\be
\frac{\partial {a}}{\partial \ln Q^2} =
{\beta}^{(N)}({a}) 
= - {\beta_0} {a}^2 
\left( 1 + \sum_{j=1}^N {c_j} {a}^j \right).
\label{betaN}
\ee
This was then modified, 
${\beta}^{(N)}({a}) \mapsto
{\widetilde{\beta}}^{(N)}({a})$, 
by fulfilling three main conditions:

1.) ${\widetilde{\beta}}^{(N)}({a})$ has the same
expansion in powers of ${a}$ as 
${\beta}^{(N)}({a})$;

2.) ${\widetilde{\beta}}^{(N)}({a}) \sim 
- {\zeta} {a}^{p}$ with
${\zeta} > 0$ and $p \leq 1$, for $a \gg 1$, in order to ensure
the absence of {Landau} singularities;

3.) ${\widetilde{\beta}}^{(N)}({a})$ is {analytic} 
function at ${a}=0$, in order to ensure 
$|{a}(Q^2) - {a}_{\rm pt}(Q^2)| < 
( {\Lambda}^2/Q^2 )^{{k}}$
for any ${k} > 0$ at large $Q^2$  (thus respecting
the {ITEP-OPE} approach).

This modification was performed by the substitution
$a \mapsto u(a) \equiv a/(1\!+\!{\eta} a)$,
${\eta} > 0$ being a parameter, and
\ba
{\widetilde{\beta}}^{(N)}({a}) & = &
- {\beta_0} \left[ 
{\kappa}( a\!-\!u(a) )\!+\!\sum_{j=0}^N 
{\widetilde{c}_j} u(a)^{j+2} \right],
\label{tbetaN}
\ea
and ${\widetilde{c}_j}$ are adjusted so that the first condition is
fulfilled
\bep
{\widetilde{c}_0} =  1\!-\!{\eta \kappa},
\quad
{\widetilde{c}_1} =  {c_1}\!+\!2 {\eta}\!-\!{\eta}^2 {\kappa}, 
\quad \rm etc.
\eep 
This procedure results in an analytic coupling
$a(Q^2)$, with $p=1$ and ${\zeta}=\beta_0 \kappa$,
and with two positive adjustable parameters $\kappa$ and $\eta$.
The QCD parameter $\Lambda$ was taken the same as in the pQCD.
Evaluation of observables
was carried out in terms of power expansion,
with the replacement ${a}^n_{\rm pt} \mapsto {a}^n$.
Further, the couplings in this set are IR infinite:
${a}(Q^2) \sim 1/(Q^2)^{{\beta_0 \kappa}} \to \infty$
when $Q^2 \to 0$.
These new $a(Q^2)$'s are analytic ($a \equiv \A_1$). 
The RScl and RSch sensitivity of the modified TPS's of
space-like observables turned out to be reduced. 
The author of Refs.~\cite{Raczka}
chose ${\kappa} = 1/{\beta_0}$; 
by fitting the predicted values of the static
interquark potential to lattice results, he obtained 
${\eta} \approx 4.1$. 

\subsection{Set of models B}
\label{subsec:ourmod}

This is the set of models for $\A_1$ constructed in 
Ref.~\cite{Cvetic:2007ad}.
A class of {IR-finite} analytic couplings
which respect the ITEP-OPE philosophy can be constructed directly.
The proposed class of couplings has three parameters ($\eta, h_1, h_2$).
In the intermediate energy region ($Q \sim 1$ GeV),
the proposed coupling has 
low loop-level and renormalization scheme dependence.
We outline here the construction. We recall expansion (\ref{apt})
for the perturbative coupling $a(Q^2)$,
where $L=\log Q^2/{\Lambda^2}$ 
and $K_{k \ell}$ are functions of the ${\beta}$-function coefficients.
This expansion (sum) is in practice usually truncated in the
index $k$ ($k \leq k_m$).
\begin{figure}[htb] 
\centering
\epsfig{file=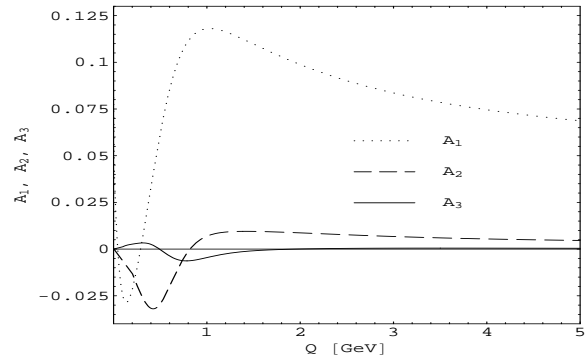,width=8.5cm,height=5.5cm}
\vspace{-0.7cm}
\caption{\footnotesize
The couplings ${\mathcal{A}_2}$ and ${\mathcal{A}_3}$,
together with the corresponding coupling ${\mathcal{A}_1}$,
are plotted as a function of $Q$, in the 
${{\overline {\rm MS}}}$-scheme, with ${\Lambda}=0.4$ GeV.
The parameters used for the couplings are 
${\eta} = 0.3$, ${h_1}=0.1$, and  ${h_2}=0$.
Figure from: Ref.~\cite{Cvetic:2007ad}.}
\label{ourmodel}
\end{figure}
The proposed coupling is obtained by modifying (the nonanalytic)
$L$'s to analytic quantities $L_0$ and $L_1$
that fall faster than any inverse power of $Q^2$
at large $Q^2$, and by adding to the truncated sum another 
quantity with such properties:
\begin{equation}
{\mathcal{A}_1}^{(k_m)}(Q^2)=
\sum_{k=1}^{k_m} \sum_{\ell=0}^{k-1} K_{l \ell}\,
\frac{(\log L_1)^{\ell}}{ L_0^k}
+ e^{-{\eta} \sqrt{x}}\, f(x),
\label{coupling}
\end{equation}
where $x = Q^2/\Lambda^2$.
The second term is only relevant in the 
{IR} region, and the first term (double sum) plays, 
in the {UV} region,
the role of the perturbative coupling.
${L_0}$ and ${L_1}$ are analytic and 
chosen aiming at a low $k_m$-dependence in the IR region.
\begin{equation}
\frac{1}{{L_i}}=\frac{1}{{L}}+
\frac{\; e^{{\nu_i}(1-\sqrt{x})}}{1-x} {g_i}(x),  \quad 
{\nu_i}>0,  \quad i=0, 1.
\label{Ldef}
\end{equation}
Functions ${g_i(x)}$ are chosen in simple meromorphic form
\begin{eqnarray}
  {g_0}(x) &=& \frac{2 x}{(1+{\nu_0})+x(1-{\nu_0})},\quad 0<{\nu_0}<1; \\
  {g_1}(x) &=& \frac{{d} e^{-{\nu_1}}+x ({d}+1-d e^{-{\nu_1}})}{{d} + x},\quad 
{d}>0,
\end{eqnarray}
with the constants fixed at typical values ${\nu_0}=1/2$ and ${\nu_1}=
{d}=2$. The additional expoinential term in (\ref{coupling})
is chosen in a similar meromorphic form
\begin{equation}
e^{-{\eta} \sqrt{x}}\, f(x)=
{h_1} \, \frac{1+{h_2}\, x}{(1+x/2)^2}\, e^{-{\eta} \sqrt{x}},
\label{couplingIR}
\end{equation}
Results for $\A_1$, $\A_2$ and $\A_3$, 
for specific typical values of parameters
$\eta$, $h_1$ and $h_2$, are shown in Fig.~\ref{ourmodel}.
Couplings $\A_2$ and $\A_3$ are constructed
via $\tA_2$ and $\tA_3$,
according to the procedure described in Sec.~\ref{sec:hipow},
Eqs.~(\ref{A2A3}).

A general remark: if $\A_1(Q^2)$ differs from the
perturbative $a(Q^2)$ by less than any negative power of $Q^2$
at large $Q^2$ ($\gg \Lambda^2$), then the same is true for the
difference between any $\tA_k(Q^2)$ and ${\widetilde a}_k(Q^2)$
($k=2,3,\ldots$).

\section{Summary}
\label{sec:summ}

Various analytic (anQCD) models, i.e., 
analytic couplings ${{\A_1}(Q^2)}$,
were reviewed, including some of those beyond the 
minimal analytization (MA) procedure.

Analytization of the higher powers
${a^n \mapsto  \A_n}$ was considered;
an RGE-motivated approach, 
which is applicable to any model of 
analytic ${\A_1}$, was described.
Analytization of noninteger powers $a^\nu$
in MA model was outlined.

Evaluation methods for space-like and time-like
observables in anQCD models were reviewed.
A large-${\beta_0}$-motivated expansion of 
space-like inclusive observables
is proposed, with the resummed leading-${\beta_0}$ part;
on its basis, an evaluation
of such observables in anQCD models is proposed: 
truncated analytic series (TAS).
Several evaluated observables in various 
anQCD models were 
compared to the experimental data.
We recall that evaluated expressions
for space-like observables in anQCD respect the physical
analyticity requirement even at low energy, in contrast to those
in perturbative QCD (pQCD).

Finally, specific classes of analytic couplings 
${\A_1}(Q^2)$ which 
preserve the OPE-ITEP philosophy were discussed, 
i.e., at high $Q^2$ they approach the pQCD coupling
faster than any inverse power of $Q^2$.
Such analytic couplings should eventually
enable us to use the OPE approach in anQCD models.



\noindent
{\bf Acknowledgments}

This work was supported in part by Fondecyt (Chile) 
Grant No.~1050512 (G.C.)
and by Conicyt (Chile) 
Bicentenario Project PBCT PSD73 (C.V.).
This work is partly based on a talk given 
by one of us (G.C.) at II Latin American Workshop on 
High Energy Phenomenology (II LAWHEP), 
S\~ao Miguel das Miss\~oes, RS, Brazil, December 3-7, 2007.

\end{document}